\documentclass[twocolumn, letter]{jpsj3}

\usepackage[T1]{fontenc}
\usepackage{textcomp}
\usepackage{lmodern}
\def\journal#1#2#3#4#5{#1: #2 {\bf #3} (#5) #4.}

\def\arXiv#1#2{#1: cond-mat/#2.}
\def\book#1#2#3{#1: \textit{#2}, #3.}
\def\JPSJ{J.\ Phys.\ Soc.\ Jpn.}
\def\PR{Phys.\ Rev.}
\def\PRL{Phys.\ Rev.\ Lett.}
\def\PRB{Phys.\ Rev.\ B}

\def\RMP{Rev.\ Mod.\ Phys.}

\def\JPC{J. Phys. C}

\newcommand{\Vg}{\ensuremath{V_\mathrm{g}}}

\newcommand{\EF}{\ensuremath{E_\mathrm{F}}}
\newcommand{\kB}{\ensuremath{k_\mathrm{B}}}
\newcommand{\Eg}{\ensuremath{E_\mathrm{g}}}



\def\runauthor{\textsc{H. Shinaoka} \textit{et al.}}
\makeatletter
\def\@typeset{}
\makeatother

\begin{document}

\title{Theory of Electron Transport near Anderson-Mott Transitions}
\author{Hiroshi {\sc SHINAOKA}$^{1}$\thanks{E-mail address: h.shinaoka@aist.go.jp}\thanks{Present address: Nanosystem Research Institute, AIST, Tsukuba 305-8568}  and Masatoshi {\sc IMADA}$^{2,3}$}
\inst{
$^{1}$ Institute for Solid State Physics, University of Tokyo, Kashiwanoha, Kashiwa, Chiba, 277-8581\\
$^{2}$ CREST, JST, 7-3-1 Hongo, Bunkyo-ku, Tokyo 113-8656\\
$^{3}$ Department of Applied Physics, University of Tokyo, Hongo, Bunkyo-ku, Tokyo 113-8656} 
\recdate{\today}

\date{\today}
\abst{
We present a theory of the DC electron transport in insulators near Anderson-Mott transitions under the influence of coexisting electron correlation and randomness.
At sufficiently low temperatures, the DC electron transport in Anderson-Mott insulators is determined by the single-particle density of states (DOS) near the Fermi energy ($\EF$).
Anderson insulators, caused by randomness, are characterized by a nonzero DOS at $\EF$.
However, recently, the authors proposed that coexisting randomness and short-ranged interaction in insulators open a \textit{soft Hubbard gap} in the DOS, and the DOS vanishes only at $\EF$.
Based on the picture of the soft Hubbard gap, we derive a formula for the critical behavior for the temperature dependence of the DC resistivity.
Comparisons of the present theory with experimental results of electrostatic carrier doping into an organic conductor $\kappa$-(BEDT-TTF)$_2$Cu[N(CN)$_2$]Br demonstrate the evidence for the present soft-Hubbard scaling.
}
\kword{Mott transition, electron correlation, Anderson localization, randomness, disorder, Anderson-Hubbard model, single-particle density of states, soft gap}

\maketitle

Metal-insulator transition (MIT) has been one of the central issues in condensed matter physics~\cite{Imada98}.
When the electron correlation drives a MIT, e.g. as a Mott transition~\cite{Mott49}, it opens a gap in the single-particle DOS.
On the other hand, Anderson transitions induced by randomness are characterized not by a vanishing carrier number but by a vanishing relaxation time~\cite{Anderson58}.
Therefore, the Anderson insulators exhibit no gap in the single-particle DOS.

\begin{figure}[!]
 \centering
 \includegraphics[width=.4\textwidth,clip]{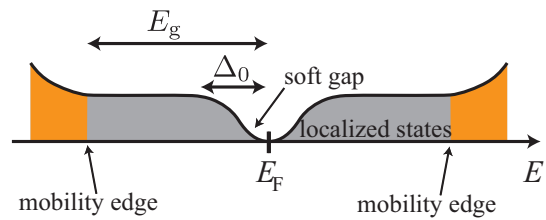}
 \caption{(Color online) Schematic picture of single-particle DOS in Anderson-Mott insulator.}
 \label{fig:gap}
\end{figure}

In real materials, however, electron correlation and randomness coexist inevitably.
Let us consider how a single-particle gap behaves when randomness (e.g., spatially fluctuating potentials) is introduced into a Mott insulator.
One might think that, when the disorder is sufficiently strong, localized impurity levels fill up and collapse the gap completely.
This simple expectation, however, is not correct; electron correlations open a \textit{soft gap} in the Anderson insulator
(see Fig.~\ref{fig:gap}).

A Coulomb gap proposed by Efros and Shklovskii~\cite{Efros75, Shklovskii84} (ES) is a typical example of such correlation-induced soft gaps.
They showed that, assuming a nonzero DOS at the $\EF$, the ground state is unstable against a particle-hole excitation under the influence of the long-range Coulomb interaction.
As a result, the DOS, $A$, vanishes toward $\EF$ as
\begin{eqnarray}
	A(E) &\propto& |E-\EF|^{d-1},
\end{eqnarray}
where $E$ is energy and $d$ is the spatial dimension.
Since the ES mechanism is based on the excitonic effects, the Coulomb gap vanishes when the short-range part becomes dominant, e.g., near MITs where the dielectric constant diverges.

On the other hand, recently, the authors found an unconventional soft gap with the help of numerical analyses on the Anderson-Hubbard model with coexisting on-site repulsions and diagonal disorders within the Hartree-Fock approximation~\cite{Shinaoka09a, Shinaoka09b}.
Even though only the short-range interaction is present in this model, we observed the formation of a \textit{soft Hubbard (SH) gap} for $d=1,2$, and $3$. We found that the DOS is scaled in $E$ near $\EF$ as
\begin{equation}
	A(E) = \alpha \exp\left[{- \left\{-\gamma\log \left|E-\EF\right|\right\}^d}\right],\label{eq:DOS-SH-old}
\end{equation}
where $\alpha$, and $\gamma$ are positive constants.
To clarify the origin of the SH gap, we assumed the existence of low-energy multiply-excited states with electronic structures that are globally relaxed from those of the ground state.
Indeed, by considering the ground-state stability against these multiply excited states, we analytically reproduced the numerically observed  DOS[eq.~(\ref{eq:DOS-SH-old})] successfully.
Such low-energy multiply-excited states are characteristic to random systems that have multivalley energy landscape or many (nearly-)degenerate metastable states.
Indeed, the insulator with the SH gap is characterized by a nonzero spin-glass (Edwards-Anderson) order parameter~\cite{Tusch93, Shinaoka10a}, in contrast to the uncorrelated Anderson insulators; there exist many excited states nearly degenerated with the ground state generically in spin/charge-glass phases.

The long-range part of the Coulomb interaction is suppressed to low energies near MITs. Therefore the effects of the short-range part practically determines the electronic structures, and the SH mechanism becomes dominant in the experimentally accessible energy scale.
The DC transport measurement is a useful and effective tool for investigating the DOS at low energies with a high resolution in insulating phases.
In the previous study~\cite{Shinaoka09b}, we derived the temperature ($T$) dependence of the DC resistivity in the presence of the SH gap.
Indeed, we found that experimental results for SrRu$_{1-x}$Ti$_x$O$_3$ are consistent with the SH scaling near the MIT.
Recent development of electrostatic carrier doping using the field-effect transistor principle now allows us fine tuning of MITs in strongly correlated electronic systems.
This motivates us to further investigate the DC electron transport in Anderson-Mott insulators, in particular, the critical behavior of the DC resistivity. 
%

In this letter, we clarify the electron transport near the Anderson-Mott transition in the presence of the SH gap. Particularly, we investigate the critical behavior of the $T$ dependence of the DC resistivity.
We compare the present theory with a recent experimental result of the electrostatic carrier doping into a thin-film single crystal of an organic superconductor $\kappa$-(BEDT-TTF)$_2$Cu[N(CN)$_2$]Br (, hereafter abbreviated as $\kappa$-Br)~\cite{Kawasugi09}.

Figure~\ref{fig:gap} illustrates the single-particle DOS in an Anderson-Mott insulator. For simplicity, we consider the electron-hole symmetric case here. High-energy extended states are separated from the low-energy localized states by a mobility edge located at the energy distance  $E_\text{g}$ from $\EF$.  The MIT is identified by a vanishing $E_\text{g}$ ,which corresponds to a divergence of the localization length $\xi$ at $\EF$.

At high $T$ i.e. $\kB T \gg \Eg$, carriers are thermally excited over the gap between the mobility edges, and they dominate the conduction. Thus the $T$ dependence of the DC resistivity follows the Arrhenius law as
\begin{eqnarray}
 \rho(T) &=& \rho_0 \exp\left( \frac{\Eg}{\kB T}\right).
\end{eqnarray}
On the other hand, at low $T$ compared to $\Eg$, the electron hopping between localized states near $\EF$ becomes dominating because the density of the thermally excited carriers vanishes exponentially as $T$ decreases for $\kB T < \Eg$.

Mott showed that, when there is a non-vanishing DOS at $\EF$, the conduction is dominated by hoppings between localized states within a narrow energy centered at $\EF$ at sufficiently low $T$ [variable-range hopping (VRH)]. In the VRH, the $T$ dependence of the DC resistivity exhibits a universal behavior given by
\begin{eqnarray}
 \rho(T) &=& \rho_0 \exp\left[\left(\frac{E_0}{\kB T} \right)^{1/(d+1)}\right]. \label{eq:VRH}
\end{eqnarray}
Here $E_0$ is an energy scale set by $\xi$ and $A(\EF)$.
Because the VRH explicitly depends on the DOS near $E_\mathrm{F}$, the formation of the SH gap modifies the $T$ dependence of the resistivity given by eq.~(\ref{eq:VRH}) qualitatively.

We start with a brief review of the phenomenological theory of the SH gap following refs.~\citen{Shinaoka09a} and \citen{Shinaoka09b}, and extend it for the later use.
%
Now we consider a single-particle excitation from the ground state by adding an electron at a certain position/site.
The SH theory claims that the excitation energy has a lower bound $\Delta~(>0)$ arising from the ground-state stability against multiply excited states. Note that, in uncorrelated Anderson insulators, $\Delta$ is always zero, leading to a non-vanishing DOS at $\EF$.
Because $\Delta$ is determined by the energy scale of electron correlations between the added electron and surrounding electrons in the ground state, $\Delta$ is given by 
\begin{eqnarray}
	\Delta &=& \Delta_0 \exp\left( - b R\right).\label{eq:DR}
\end{eqnarray}
Here, $b\propto \xi^{-1}$ and $R$ is the distance between the added electron and the electron nearest to it in the ground state.
Within the single-particle picture, the wave function of an localized electron, $\phi$, is scaled in distance from the localization center $r$ as
\begin{eqnarray}
	\phi(r) &\propto& \xi^{-d/2} \exp(-r/\xi),
\end{eqnarray}
where $\xi$ is the localization length.
Therefore, $\Delta$ is scaled in the mutual distance $R$ as
\begin{eqnarray}
	\Delta &\propto& \mid \phi(R)\mid^2 \propto \xi^{-d} \exp(-2R/\xi).\label{eq:Delta}
\end{eqnarray}
By comparing eq.~(\ref{eq:DR}) and eq.~(\ref{eq:Delta}), we obtain 
\begin{eqnarray}
	\Delta_0 &\propto& \xi^{-d}.
\end{eqnarray}

On the other hand, the distribution of $R$ 
is given by
\begin{eqnarray}
	P(R) &=& a^\prime \exp\left( - b^\prime R^d\right), \label{eq:PR}
\end{eqnarray}
where $a^\prime$ and $b^\prime$ are non-universal positive constants depending on the electron filling and $\xi$.
By integrating the distribution of $\Delta$ with respect to energy, which is derived from eqs.~(\ref{eq:DR}) and (\ref{eq:PR}), we obtain
\begin{equation}
 A(E) = \alpha \exp\left[{- \left\{-\gamma\log \left|\frac{E-\EF}{\Delta_0}\right|\right\}^d}\right]\label{eq:DOS-SH}
\end{equation}
for $|E-\EF| < \Delta_0$. Here we take into account logarithmic corrections ignored in the previous studies~\cite{Shinaoka09a, Shinaoka09b}, which is important in considering the electron transport near the MIT.
Here $\alpha$, and $\gamma(={b^\prime}^{1/d}/b)$ are positive constants.
Near the MIT, $\gamma$ is scaled with $\xi$ as $\gamma \propto \xi$, while $\alpha$ is non-critical.

Following the discussion in ref.~\citen{Shinaoka09b} and taking into account logarithmic corrections omitted in the previous studies, we obtain the $T$ dependence of the DC resistivity as
\begin{equation}
 \rho=\rho_0 \exp\left\{ c_0 \frac{\exp[-c_1| \log(k_\mathrm{B} T/\Delta_0)|^{1/d}]}{k_\mathrm{B} T}\right\},\label{eq:rho-SH}
\end{equation}
where $\rho_0$, $c_0$ and $c_1$ are positive constants. This SH scaling indicates that the DC resistivity diverges toward zero $T$ slightly slower than the Arrhenius law; it reduces to the Arrhenius law when $c_1=0$. Near the MIT, $c_0$ and $c_1$ are scaled with $\xi$ as
\begin{eqnarray}
	c_0 &=& \Delta_0 \left[1 + \frac{2}{\xi (2\alpha \Delta_0)^{1/d}}\right] \propto \xi^{-d} \rightarrow 0, \label{eq:cb-c0}\\
 c_1 &=& d^{1/d} \gamma^{-1} \propto \xi^{-1} \rightarrow 0.\label{eq:cb-c1}
\end{eqnarray}

Note that, even near the MIT where the dielectric constant diverges, the electron interactions remain long ranged at asymptotically long distances and low energies.
Thus we expect a crossover from the SH scaling to the ES scaling with decreasing $T$.
Below the crossover temperature, the DC resistivity follows the ES scaling~\cite{Efros75} as
\begin{eqnarray}
 \rho &=& \rho_0 \exp\left[\left(\frac{T_0}{T} \right)^{1/2}\right].\label{eq:rho-ES}
\end{eqnarray}

\begin{figure}[h]
 \centering
 \includegraphics[width=.4\textwidth,clip]{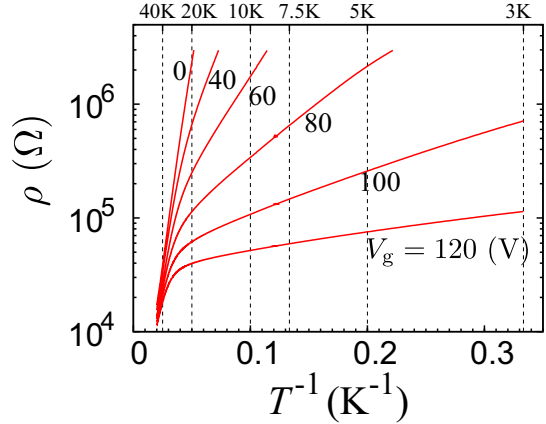}
 \caption{(Color online) Arrhenius plot of experimental data for $\Vg=0, 40, 60, 80, 100, 120$ V taken from Ref.~\citen{Kawasugi09}.}
 \label{fig:Arrhenius-Vg-all}
\end{figure}
\begin{figure}[h]
 \centering
 \includegraphics[width=.4\textwidth,clip]{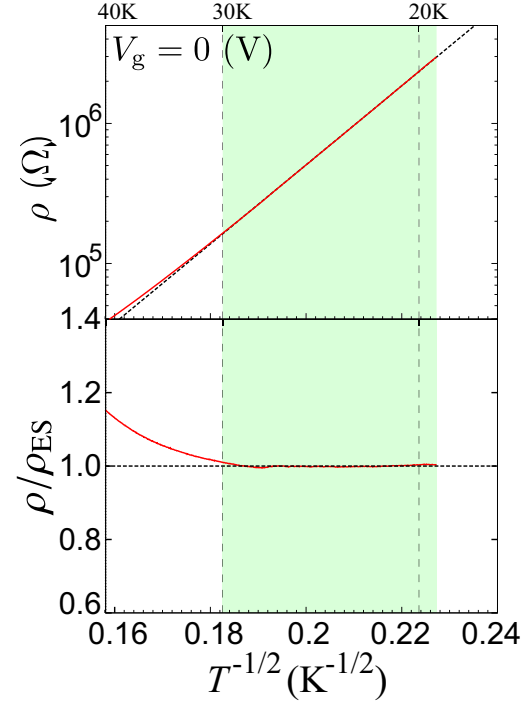}
 \caption{(Color online) Enlarged plot of Fig.~\ref{fig:Arrhenius-Vg-all} for $\Vg = 0$. The broken (black) line in (a) is a fit by the Arrhenius law, while that in (b) is a fit by the ES scaling. The lower panels in (a) and (b) show the experimental data divided by the fitted curves for the Arrhenius law and the ES scaling, respectively.}
 \label{fig:Vg-0}
\end{figure}
\begin{figure}[h]
 \centering
 \includegraphics[width=.4\textwidth,clip]{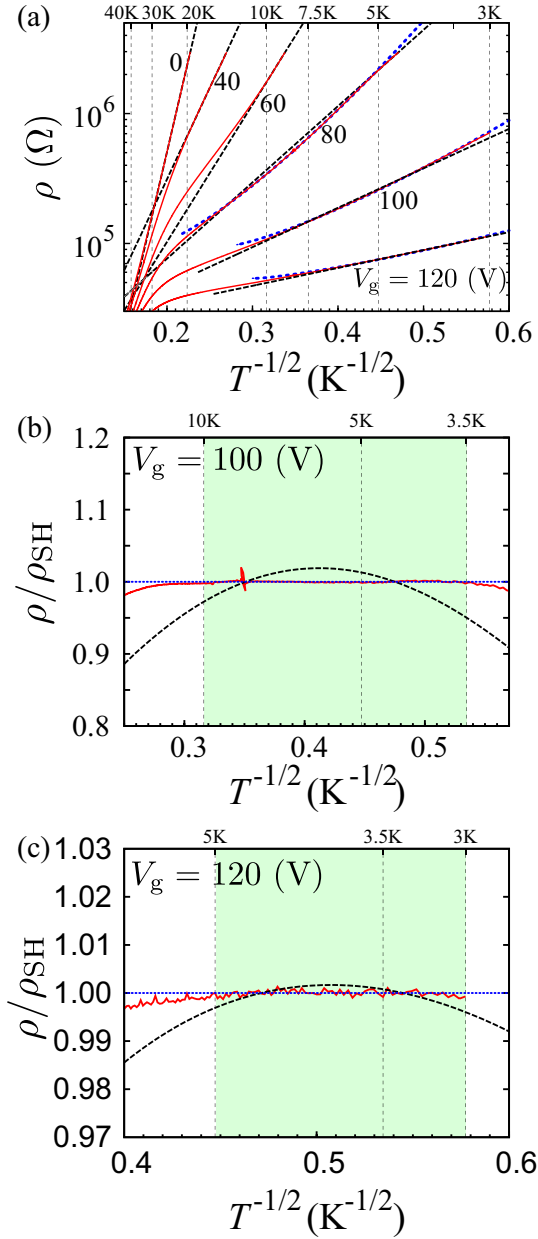}
 \caption{(Color online) (a) ES plot of experimental data (solid (red) curves) for $\Vg=0, 40, 60, 80, 100, 120$ V. The broken (black) lines are fits by the ES scaling, and the dotted (blue) lines are fit by the SH scaling. The experimental data divided by the fitted curve for the SH scaling are shown for $\Vg=100, 120$ V in (b) and (c), respectively.}
 \label{fig:ES-Vg-all}
\end{figure}
Recently, it was reported that a thin-film single crystal of an organic superconductor $\kappa$-Br adsorbed to a Si substrate becomes a Mott insulator due to a negative pressure imposed by the incompressible Si substrate~\cite{Kawasugi08}; the substrate prevents the organic crystal from thermal shrinkage at low $T$. Subsequently, Kawasugi \textit{et al.} reported successful doping of electrostatic carriers into the organic conductor by applying a gate-bias voltage~\cite{Kawasugi09}. Figure~\ref{fig:Arrhenius-Vg-all} shows the experimental data of the DC resistivity at $\Vg = 0, 40, 60, 80, 100$, and $120$ V taken from ref.~\citen{Kawasugi09}. With increasing gate-bias voltage, the divergence of the DC resistivity at low $T$ becomes weaker and weaker, indicating that the system approaches the MIT.
This fine tuning provides us with a suitable system to test the present theory.

Figure~\ref{fig:Vg-0}(a) shows an Arrhenius plot of the experimental data for $\Vg=0$ V. 
The Arrhenius law fits well with the experimental data for $T<30$ K at a first glance in the upper panel.
However, on closer inspection in the lower panel, we find a clear deviation from the Arrhenius law; the data are convex upward for $T<40$ K in the Arrhenius plot, indicating that the DC resistivity diverges toward low $T$ in a rate slower than the Arrhenius law.
Indeed, as demonstrated in Fig.~\ref{fig:Vg-0}(b), the data exhibits a perfect fit with the ES scaling rather than the Arrhenius law.
This indicates a formation of localized states near $\EF$ and the existence of a Coulomb gap under the influence of the long-ranged part of the Coulomb interaction.
Thus the insulator at $\Vg=0$ V can be regarded as a Anderson-Mott insulator rather than a \textit{pure} Mott insulator.
The randomness in $\kappa$-Br may originate from motions of the ethylene moiety in the BEDT-TTF molecule~\cite{Akutsu99, Akutsu00, Muller02, Yoneyama04}.

Now we discuss the electron transport near the MIT, e.g., at $\Vg>0$.
Figure~\ref{fig:ES-Vg-all}(a) shows an ES plot for $\Vg=0, 40, 60, 80, 100$, and $120$ V.
As increasing $\Vg$ i.e. approaching the MIT, the region where the data fit the ES scaling (ES region) is restricted to lower and lower $T$ as shown by broken (black) lines for $\Vg = 0, 40, 60$, and $80$ V.
Particularly, for $\Vg \ge 100$ V, the experimental data become concave downward down to the lowest available $T$.
This behavior is consistent with the expectation that the effects of the long-ranged part of the Coulomb interaction are suppressed to lower and lower $T$ near the MIT.
Indeed, at a higher $T$ above the ES region, as is shown by the dotted lines in Figs.~\ref{fig:ES-Vg-all}(b) and (c), the SH scaling fits the experimental data much better for $\Vg \ge 80$ V.
In the SH fit, it is numerically difficult to unambiguously determine all the four parameters of the SH scaling.
Therefore we assume $\Delta_0 = 60, 50$, and $40$ K for $\Vg = 80, 100$, and $120$ V, respectively, so that $\Delta_0$ decreases with increasing $\Vg$.
It should be noted that the fitting quality does not strongly depend on the choice of $\Delta_0$.

The parameters of the SH scaling obtained by the fitting are summarized in Table~\ref{table:param}.
The $\Vg$ dependences of the fitting parameters ($c_0$, $c_1$) do not contradict the predicted critical behavior [eqs.~(\ref{eq:cb-c0}) and (\ref{eq:cb-c1})], while the error bars are too large to identify the critical exponents.
\begin{table}[t]
		\centering
			\begin{tabular}{r|cccc}
						\hline
		$V_\text{g}$ (V)    &     $\Delta_0$ (K) &        $\rho_0$      &      $\log(c_0)$     &      $c_1$\\ \hline
		80                  &     60             &        $9.7\pm 0.1$  &      $4.8\pm 0.1$   &      $0.95\pm 0.05$\\
		100                 &     50             &        $10.1\pm 0.02$  &      $4.0\pm 0.1$   &      $0.95\pm 0.02$\\
		120                 &     40             &       $9\pm 1$  &      $4.6\pm 1.8$   &      $1.7\pm 0.8$\\
		\hline
	\end{tabular}
	\caption{Fitting parameters for the SH scaling.}
	\label{table:param}
\end{table}

In summary, we have investigated the DC electron transport in the insulators near the Anderson-Mott transition.
We have derived the formula for the critical behavior of the $T$ dependence of the DC resistivity based on the picture of the SH gap.
The present theory was tested in detail against the recent experimental result on the electrostatic carrier doping into a thin-film single crystal of $\kappa$-Br.
Indeed, we have found that the experimental result of the DC resistivity follows the SH scaling near the MIT, supporting the present theory.

A direct observation of the Coulomb gap was reported by using electron tunneling spectroscopy in Si:B~\cite{Massey95}.
Electron-tunneling measurements allow us to further investigate the critical behavior of the DOS, when they are combined with recently developed experimental techniques for the fine tuning of the MIT (e.g., the electrostatic carrier doping and X-ray irradiation in organic compounds~\cite{Sasaki07, Sasaki08, Sano10}).

\acknowledgements{
We thank Y. Kawasugi for providing us with the experimental data. We also thank H. M. Yamamoto and T. Sasaki for fruitful discussions.
}

\end{document}